# Statistical mechanics unifies different ecological patterns

Roderick C. Dewar[a,*], Annabel Porté[b]

[a]*Laboratory of Functional Ecology and Environmental Physics, INRA Centre de Bordeaux-Aquitaine,*

*B.P. 81, 33883 Villenave d'Ornon CEDEX, France*

[b]*Laboratory of Functional Ecology and Environmental Physics, INRA Centre de Bordeaux-Aquitaine,*

*33612 Cestas CEDEX, France*

\* Corresponding author at: Laboratory of Functional Ecology and Environmental Physics, INRA Centre de Bordeaux-Aquitaine, B.P. 81, 33883 Villenave d'Ornon CEDEX, France.

Tel: +33 5 57 12 24 16; fax: +33 5 57 12 24 20.

*Email address:* dewar@bordeaux.inra.fr.

---

**Abstract**

Recently there has been growing interest in the use of Maximum Relative Entropy (MaxREnt) as a tool for statistical inference in ecology. In contrast, here we propose MaxREnt as a tool for applying statistical mechanics to ecology. We use MaxREnt to explain and predict species abundance patterns in ecological communities in terms of the most probable behaviour under given environmental constraints, in the same way that statistical mechanics explains and predicts the behaviour of thermodynamic systems. We show that MaxREnt unifies a number of different ecological patterns: (i) at relatively local scales a unimodal biodiversity-productivity relationship is predicted in good agreement with published data on grassland communities, (ii) the predicted relative frequency of rare *vs.* abundant species is very similar to the empirical lognormal distribution, (iii) both neutral and non-neutral species abundance patterns are explained, (iv) on larger scales a monotonic biodiversity-productivity relationship is predicted in agreement with the species-energy law, (v) energetic equivalence and power-law self-thinning behaviour are predicted in resource-rich communities. We identify mathematical similarities between these ecological patterns and the behaviour of thermodynamic systems, and conclude that the explanation of ecological patterns is not unique to ecology but rather reflects the generic statistical behaviour of complex systems with many degrees of freedom under very general types of environmental constraints.

---

## 1. Introduction

A central goal of ecology is to understand ecological patterns in multi-species communities. While many explanations already exist for individual patterns, what is lacking is a theory which unifies and reconciles these often contrasting patterns within a common logical framework. For example, empirical studies of the relationship between community diversity and resource use show two contrasting patterns at different spatial scales. At a relatively local scale the relationship between diversity and resource use (or some proxy of resource use such as productivity) tends to be unimodal, with maximum diversity at intermediate resource levels (e.g. Al-Mufti et al., 1977; Chase and Leibold, 2002; Harpole and Tilman, 2006; Kassen et al., 2000; Rosenzweig, 1995). At regional and continental scales, however, diversity tends to increase monotonically with resource use or productivity (e.g. Currie, 1991; Currie and Paquin, 1987; Waide et al., 1999).

The mechanisms underlying these contrasting patterns remain unclear. Various explanations have been proposed involving niche competition within heterogeneous environments at relatively small scales, and, at larger scales, the influence of energy input on population size or the influence of area on geographical-range size (Gaston, 2000; Kassen et al., 2000; Wright, 1983). According to Gaston (2000), 'Reconciliation of the patterns in biodiversity that are observed at different scales may provide significant insights into their determinants'.

Another area where reconciliation might provide new insights is the current debate concerning the mechanisms of community assembly. Evidence has been presented for both 'neutral' stochastic processes (no species differences in per capita birth, death and dispersal rates) (Hubbell, 2001; Volkov et al. 2003, 2005) and 'non-neutral' deterministic processes (competition among ecologically different species) (Harpole and Tilman, 2006). More generally, a theory that unifies these contrasting patterns might also lead to a common understanding of other ecological patterns such as the species-energy power law (e.g. Currie et al., 2004; Wright, 1983), energetic equivalence (e.g. Allen et al., 2002; Damuth, 1987; Enquist et al., 1998) and the self-thinning power law (e.g. Dewar, 1993, 1999; Enquist et al., 1998). Finally, new fundamental insights are also needed to answer the outstanding question of just why there are so many species on Earth (e.g. Tilman, 2000).

In this paper we show that statistical mechanics explains and reconciles these disparate ecological patterns, thus providing a unifying framework for quantitative ecology. Statistical mechanics, first developed by Maxwell (1871), Boltzmann (1898) and Gibbs (1902), aims to understand and predict from a probabilistic viewpoint the macroscopic behaviour of complex systems consisting of large numbers of interacting microscopic components. Typically, due to the large number of components, the same macroscopic behaviour can be realized in many different ways microscopically. Statistical mechanics predicts the most probable macroscopic behaviour, *i.e.* the one that can be realized in the greatest number of ways microscopically. The relevance of statistical mechanics to ecological communities stems from the large number of 'microscopic components', namely, the individuals of different species. This opens the way to a statistical explanation of ecological patterns: they are expressions of the community-level behaviour that can be realised in the greatest number of ways at the individual level.

Such a statistical explanation of ecological patterns contrasts with deterministic explanations in terms of the underlying population dynamics (e.g. Hubbell, 2001; MacArthur, 1972; Tilman, 1982). Historically, we may draw a parallel here with the behaviour of thermodynamic systems – including the famous $2^{nd}$ law (entropy increases) – for which a dynamical explanation was also sought, notably by Boltzmann (1898) himself in an argument since refuted by others (e.g. Jaynes, 1971). Maxwell (1871) was the first to realise that the $2^{nd}$ law is in fact a statistical law, not a dynamical one, and that the apparently deterministic behaviour of macroscopic systems reflected the fact that it was the most probable behaviour by an exceedingly large margin. More recent theoretical and experimental studies have confirmed this point (Evans and Searles, 2002; Wang et al., 2002). In this paper we show that the statistical explanation of ecological patterns is not merely analogous to that of thermodynamic behaviour; mathematical similarities between the two explanations point to their common origin in the generic statistical behaviour of complex systems under very general types of environmental constraints.

Mathematically, our application of statistical mechanics to ecological communities consists of the maximisation of relative entropy (MaxREnt). The concept of relative entropy was originally introduced in the guise of its negative, the Kullback-Leibler divergence, as a measure of the difference between two probability distributions (Kullback and Leibler, 1951). Then, in the context of statistical inference, the MaxREnt principle (also known as the Principle of Minimum Discrimination Information, and the Principle of Minimum Cross-Entropy) was proposed as a method of updating a prior distribution to a posterior distribution in the light of new data (Kullback, 1968). As a generalisation of another inference principle, maximum Shannon entropy (MaxEnt), MaxREnt sits more naturally within the framework of Bayesian probability theory (Jaynes, 2003). In particular, there has been growing interest in the use of MaxREnt as a tool for statistical inference in ecology (e.g. Burnham and Anderson, 2002; Clark and Gelfand, 2006).

Historically, however, the origins of MaxEnt lie not in statistical inference but in statistical mechanics (Gibbs, 1902; Jaynes 1957, 1978), as a theoretical tool to explain and predict the macroscopic behaviour of complex systems in terms of their most probable behaviour. It is natural, then, to consider the more general MaxREnt principle also as a tool for statistical mechanics, not just for statistical inference. Here we demonstrate the use of MaxREnt as a statistical mechanical tool to explain and predict the behaviour of ecological communities under given environmental constraints.

In Section 2, following Jaynes (1957, 1978), we explain the physical rationale for MaxREnt as a tool for statistical mechanics (rather than statistical inference). Section 3 describes how we used MaxREnt to predict species abundance distributions in the context of a simple model of community resource use. Section 4 shows how the predictions of MaxREnt explain and unify a disparate collection of ecological patterns in multi-species communities. Section 5 highlights the mathematical similarities between these predicted ecological patterns and the behaviour of thermodynamic systems. Section 6 summarises our conclusion. Appendices A-E provide mathematical details.

## 2. Maximum relative entropy and statistical mechanics

### 2.1 MaxREnt as a tool for statistical inference

In the context of information theory, given a probability distribution $p_i$ over some set of outcomes (labelled $i$), the Shannon entropy of $p \equiv \{p_i\}$, defined by

$$H_S(p) = -\sum_i p_i \log_e p_i \,, \tag{1}$$

is a measure of the associated uncertainty or missing information about the outcomes (Jaynes, 2003; Shannon, 1948). As a tool for statistical inference, maximisation of Shannon entropy (MaxEnt) with respect to $p$, subject to known data constraints, constructs the most conservative, non-committal $p$ consistent with the data (e.g. Jaynes, 2003) in the sense that use of any other distribution would assume more information than is known from the data.

However, in the context of Bayesian inference a more general and fundamental quantity is the relative entropy between two distributions $p$ and $q$, defined by

$$H(p\|q) \equiv -\sum_i p_i \log_e \frac{p_i}{q_i} \tag{2}$$

The relative entropy satisfies Gibbs' inequality $H(p\|q) \le 0$ with equality if and only if $p = q$. The Kullback-Leibler divergence $D(p\|q) = -H(p\|q) \ge 0$ (Kullback and Leibler, 1951) is a positive measure of the difference between $p$ and $q$. In terms of information $D(p\|q)$ represents the information gained when $p$ is used instead of $q$. In Bayesian inference, the MaxREnt principle (also known as the Principle of Minimum Discrimination Information, and the Principle of Minimum Cross-Entropy) is a method of updating a prior distribution $q$ to a posterior distribution $p$ in the light of new data (Kullback, 1968): choose the $p$ that represents the smallest information gain $D(p\|q)$. MaxREnt has also been used as a theoretical basis for model selection by the Akaike Information Criterion (Akaike, 1973; Burnham and Anderson, 2002). As is evident from Eqs. (1) and (2), MaxEnt is the special case of MaxREnt when the prior distribution $q$ is uniform over the outcomes.

## 2.2 MaxREnt as a tool for statistical mechanics

Statistical mechanics aims to explain and predict the macroscopic behaviour of complex systems consisting of large numbers of microscopic degrees of freedom, which are subject to given physical constraints (experimental conditions). Generally the physical constraints ($C$) are insufficient to determine with complete certainty the microscopic state or 'microstate' $i$ of the system. A central quantity of interest, therefore, is the probability $p_i$ that the system is in microstate $i$. From $p_i$ any macroscopic quantity $Q$ is predicted as the expectation value

$$\overline{Q} = \sum_i p_i Q_i \quad . \tag{3}$$

($Q_i$ = value of $Q$ in microstate $i$). The central problem of statistical mechanics is to construct $p$.

Gibbs (1902) constructed $p$ using (what is now called) MaxEnt. That is, he maximised $H_S(p)$ (Eq. 1) with respect to each $p_i$, subject to the physical constraints $C$ on the system and normalization of $p$. Typically the physical constraints $C$ take the form of various known macroscopic averages expressed as in Eq. (3) (e.g. the average total energy in the case of a system maintained in equilibrium with a heat reservoir). MaxEnt then leads to the standard probability distributions of equilibrium statistical mechanics (e.g. Jaynes, 1957, 1978). Historically, the fact that the Boltzmann distribution, which had previously been derived by microstate counting arguments, could be rederived in this more general way was why Gibbs (1902) proposed the MaxEnt algorithm, although its physical meaning remained obscure for many years.

The physical meaning of MaxEnt was first elucidated by Jaynes (1957) (see also Jaynes, 1978). The key point is that we are concerned with predicting *via* Eq. (3) only those aspects of the behaviour of complex systems which are experimentally reproducible. If every time the same physical constraints ($C$) are applied the same macroscopic behaviour is reproduced, then knowledge of $C$ alone must be sufficient to predict that behaviour. The myriad microscopic details that differ from one repetition of the experiment to another must be irrelevant to the prediction of reproducible behaviour, and can be eliminated from the outset. Therefore, in order to predict reproducible behaviour from Eq. (3), it suffices to encode into $p$ only the information

*C*. But this is just what MaxEnt does; by maximising the missing information $H_S(p)$ with respect to *p* subject to the physical constraints *C* (and normalisation), *p* only encodes the information *C*. Any distribution other than *p* has less missing information and so encodes more information than *C*.

The underlying physical reason why reproducible behaviour emerges at all – and thus why MaxEnt works in practice – lies in the vast number of underlying microscopic components alluded to in Section 1. This underlying multiplicity ensures that each of the overwhelming majority of microstates compatible with the constraints looks the same on macroscopic scales, so that on those scales it does not matter in which of these microstates the system finds itself. This microscopic redundancy ensures that the same macroscopic behaviour is reproduced whenever the constraints are applied (Jaynes, 1957, 1978).

The same physical interpretation may be extended to MaxREnt, the Bayesian generalisation of MaxEnt. In Eq. (2), *q* represents the microstate probability distribution in the absence of the physical constraints *C*, while *p* represents the microstate probability distribution in the presence of the constraints *C*. As a statistical mechanical tool, MaxREnt constructs *p* by maximising $H(p\|q)$ (*i.e.* minimising the information gain $D(p\|q)$) with respect to each $p_i$ subject to the constraints *C* and normalisation of *p*, with *q* held fixed. The posterior distribution *p* so constructed predicts the reproducible behaviour of the system in the presence of the constraints *C*, in the more general case where the prior distribution *q* in the absence of the constraints is non-uniform. As we shall see, *q* may be non-uniform even when it describes complete ignorance about the microstates.

That experimentally reproducible behaviour is synonymous with the most likely behaviour is expressed mathematically by the fact that the MaxREnt distribution *p* is the most spread-out microstate distribution compatible with the constraints, and so *p* gives non-zero weight to the maximum number of microstates.

We emphasise that although the MaxREnt principle is the same mathematically whether it is used as a tool for statistical inference or statistical mechanics, its physical significance as a predictor of experimentally reproducible behaviour emerges only in the latter context.

## 2.3  *The prior distribution q of complete ignorance*

In our study we examined the hypothesis that detailed dynamical information (e.g. about per capita birth, death and dispersal rates) is irrelevant to predicting many of the ecological patterns observed in ecological communities, and that instead these patterns are determined by very general community-level constraints on space and resources ($C$). Therefore we made no prior assumptions whatsoever about the behaviour of the community in the absence of those constraints. The prior distribution $q$ was then determined as the distribution describing complete ignorance about the community microstates. In adopting the ignorance prior $q$, we are assuming that birth, death and dispersal rates are in reality dynamic variables which are ultimately determined by the community-level resource constraints, so that only the latter information needs to be encoded into $p$.

We obtained the ignorance prior $q$ relevant to our problem from a previously derived mathematical result (Rissanen, 1983). Our ignorance prior (Eq. 8 below) is non-uniform so that MaxREnt rather than MaxEnt is the appropriate algorithm. More generally, Jaynes (2003) has shown how the ignorance prior may be derived from the underlying symmetries of the problem (technically, a transformation group is identified under which $q$ is invariant). The ignorance prior therefore depends on the problem.

## 3.  **Methods**

### 3.1  *A simple model of community resource use*

We applied MaxREnt to a simple model of a multi-species community subjected to very general community-level constraints (Volkov et al., 2004). Specifically, we considered a community assembled from a set of $S$ possible species ($j = 1, 2 \ldots S$), which is limited by space and a single resource such as water or nitrogen. We assumed that these limitations constrain the mean total number of individuals ($\overline{N}$) and mean community-level resource use ($\overline{R}$) to adopt certain values. Thus

$$\sum_{j=1}^{S} \overline{n}_j = \overline{N} \tag{4}$$

and

$$\sum_{j=1}^{S} \overline{n}_j r_j = \overline{R} \tag{5}$$

where $n_j$ and $r_j$ are, respectively, the number of individuals and the per capita resource use of species $j$, and the overbar indicates the mean value (*i.e.* the average over some spatial and temporal scales). We assumed that each $r_j$ is a fixed constant. Different species can have different values of $r_j$, and in this sense the model can be considered to be non-neutral. However, we make no assumptions whatsoever about the underlying population dynamics, such as the species per capita rates of birth, death and dispersal, in terms of which neutrality is usually defined (Hubbell, 2001).

In general, due to the effects of species interactions on per capita resource use, each $r_j$ might depend on all the $n_k$ ($k = 1, 2 \ldots S$). In making the simplifying assumption that each $r_j$ is a fixed constant, independent of the $n_k$, we do not deny that those effects exist. Rather, we are exploring the hypothesis that those details are irrelevant for the prediction of species abundances, and that the effects of species interactions on species abundances and their correlations are largely captured through the community-level constraints, Eqs. (4) and (5).

Although in Section 4.1 we will derive some general results which are valid for arbitrary fixed $r_j$, we will illustrate these results for the particular case where the spectrum of $r_j$ values has 'ground state' $r_1 = 0$, second level $r_2 = \Delta$, and subsequent levels increasing as some power $\alpha$ of the species index $j$ up to $j = S$:

$$r_j = \Delta (j-1)^{\alpha} \qquad , \qquad j = 1 \ldots S \tag{6}$$

Biologically, Eq. (6) is motivated by various theories of metabolic scaling (Demetrius, 2006; West et al., 1997) which imply that the per capita resource use $r_j$ (a proxy for metabolic rate) and the per capita (adult) mass $m_j$ of species $j$ have a power-law relationship

$$r_j \propto m_j^{\alpha} \qquad (7)$$

If we distinguish different species $j$ according to their values of $m_j$ (i.e., introduce a small mass increment $\delta m$, let $m_j = (j - 1)\delta m$ and define species $j$ to consist of individuals whose masses lie between $m_j$ and $m_{j+1}$) then Eq. (7) implies Eq. (6), in which we can interpret $\alpha$ as the metabolic scaling exponent. A novel molecular model of metabolism in organisms predicts that $\alpha = 2/3$ for small animals and annual plants, $\alpha = 3/4$ for large animals and $\alpha = 1$ for perennial plants, consistent with the available evidence from recent empirical studies (Demetrius, 2006 and references cited therein). An alternative macroscopic model of metabolism (West et al., 1997) predicts the universal metabolic scaling exponent $\alpha = 3/4$. However, we emphasise that the key results of our study (Section 4.1) are valid for arbitrary fixed $r_j$ values.

### 3.2   Microstates and the ignorance prior q

We specified the microstate of the community by the vector $\boldsymbol{n} = (n_1, n_2 \ldots n_S)$ whose $j^{\text{th}}$ component $n_j$ is the population of species $j$.  We chose species population here rather than species biomass (Shipley et al. 2006) because our aim was to explain empirical ecological patterns that are expressed in terms of numbers of individuals. We then constructed the joint probability $p(\boldsymbol{n}) = p(n_1, n_2 \ldots n_S)$ that species 1 has population $n_1$, species 2 has population $n_2$ *etc.* by maximising the relative entropy $H(p\|q)$ (Eq. 2) subject to Eqs. (4) and (5) and normalisation of $p(\boldsymbol{n})$, where the prior distribution $q(\boldsymbol{n})$ describes complete ignorance about $\boldsymbol{n} = (n_1, n_2 \ldots n_S)$.

In the absence of any constraints, each $n_j$ can in principle take any of the values 0, 1, 2 … $\infty$ and it is the fact that $n_j$ is unbounded above which precludes the use of a uniform ignorance prior $q$. As shown in Appendix A, a previously published mathematical result (Rissanen, 1983) implies that the probability distribution describing complete uncertainty about $\boldsymbol{n}$ is approximately

$$q(\boldsymbol{n}) \equiv q(n_1, n_2 \dots n_S) = \frac{1}{n_1 + 1} \frac{1}{n_2 + 1} \cdots \frac{1}{n_S + 1} = \prod_{j=1}^{S} \frac{1}{n_j + 1} \qquad (8)$$

which gives less weight to larger $n_j$ values (the fact that this prior distribution is not normalised is discussed in Appendix A). The MaxREnt solution for $p(\boldsymbol{n})$ was then obtained using the standard method of Lagrange multipliers (Appendix B). For our simple two-constraint problem the Lagrange multipliers may be solved numerically using any convenient iterative procedure; for more complex problems a useful algorithm is described by Agmon et al. (1979).

One might object on physical or biological grounds that the $n_j$ are allowed to extend to infinity. However, to do otherwise would be to introduce information about the $n_j$ into $q(\boldsymbol{n})$ which would no longer represent an ignorance prior. The physical and biological restrictions on species abundances are incorporated into $p$ only through the resource constraints, Eqs. (4) and (5). Introducing prior ceiling values for the $n_j$ would bias $p$ and lead to predictions that reflected information other than Eqs. (4) and (5) (although the effect would be numerically small if the ceiling values were sufficiently large).

### 3.3 Species diversity and species number

The species diversity predicted by MaxREnt was quantified in terms of the Shannon diversity index (Sherwin et al., 2006; Whittaker, 1972), *i.e.* the Shannon entropy of the relative abundance distribution $\overline{n}_j / \overline{N}$ given by

$$H_n = -\sum_{j=1}^{S} \frac{\overline{n}_j}{\overline{N}} \log_e \frac{\overline{n}_j}{\overline{N}} \qquad (9)$$

The value of $H_n$ ranges from a minimum of zero (only one species present) to a maximum of $\log_e S$ (all $S$ species equally abundant). We chose to present our results in terms of the exponential of the Shannon index, $\exp(H_n)$, whose value ranges from 1 to $S$ and is therefore simpler to interpret. We used $\exp(H_n)$ to characterise the diversity–resource relationship at small scales (see Section 3.4).

As a subset of the $S$ possible species, we quantified the number of species with appreciable abundance, $S^* \leq S$, as the number of species $j$ with mean population greater than or equal to one ($\bar{n}_j \geq 1$). We used $S^*$ to characterise the species number–resource relationship at large scales (see Section 3.4).

### 3.4   Analyses at small and large scales

On small spatio-temporal scales we examined the relationship between $\exp(H_n)$ and the community resource use $\bar{R}$, holding $S$ and $\bar{N}$ at fixed finite values (Sections 4.1–4.4). We compared predicted and observed species abundances for fixed $S$ and $\bar{N}$ (Section 4.4) using published data on nitrogen-limited grasslands of the Cedar Creek Long Term Ecological Research (CDR) site, Minnesota, consisting of $S = 26$ possible species and $\bar{N} = 62$ individuals m$^{-2}$ (Harpole and Tilman, 2006; Tilman, 1984). As detailed in Appendix C, we estimated the community-level rate of soil nitrogen (N) use to be $\bar{R} = 5.9$ g N m$^{-2}$ yr$^{-1}$. In place of Eq. (6), the per capita N use of each species, $r_j$ (g N  yr$^{-1}$ individual$^{-1}$), was derived from published values of $R_j^*$ (mg N kg$^{-1}$ soil), an index of the competitiveness of species $j$ for N (Eq. C.1). We compared the predicted and observed relative abundance distributions $\bar{n}_j / \bar{N}$ at $\bar{R} = 5.9$ g N  m$^{-2}$ yr$^{-1}$, as well as the predicted and observed changes in $\bar{n}_j / \bar{N}$ when $\bar{R}$ was increased, keeping $S = 26$ and $\bar{N} = 62$ individuals m$^{-2}$ fixed.

On larger scales the spectrum of possible species from which the community is assembled can increase through speciation and immigration and is then in principle unbounded above, so we set $S = \infty$. Moreover, $\bar{N}$ may also adapt dynamically on longer time scales, so we relaxed the constraint on $\bar{N}$ (Sections 4.5–4.6). The value of $\bar{N}$ was then determined by the remaining constraint on $\bar{R}$. We then examined the relationship between $S^*$, the number of species with appreciable abundance ($\bar{n}_j \geq 1$) and $\bar{R}$ (Appendix D).

## 4.    Results

### 4.1    *General solution for arbitrary per capita resource use spectrum*

As shown in Appendix B, the MaxREnt solution for the joint distribution $p(\boldsymbol{n})$ has the multiplicative form

$$p(\boldsymbol{n}) \equiv p(n_1, n_2 \ldots n_S) = \prod_{j=1}^{S} p_j(n_j) \tag{10}$$

in which $p_j(n)$, the probability that species $j$ has population $n$, is given by

$$p_j(n) = \frac{x_j^{n+1}}{n+1} \frac{1}{\left| \log_e (1 - x_j) \right|} \qquad (j = 1, 2 \ldots S) \tag{11a}$$

where

$$x_j \equiv e^{-\beta(r_j - \mu)} \tag{11b}$$

In Eq. (11b), $\mu$ and $\beta$ parameterise the Lagrange multipliers associated with the constraint Eqs. (4) and (5) (Appendix B); they are determined numerically by adjusting their values so that Eqs. (4) and (5) are satisfied. Thus both $\mu$ and $\beta$ are functions of $\overline{N}$ and $\overline{R}$ (as well as $S$). The ecological interpretation of $\mu$ and $\beta$ is discussed in Appendix E.

The mean abundance of species $j$ is the expectation value of $n$ over $p_j(n)$:

$$\overline{n}_j = \sum_{n=0}^{\infty} n p_j(n) = \frac{x_j}{1 - x_j} \frac{1}{\left| \log_e (1 - x_j) \right|} - 1 \qquad (j = 1, 2 \ldots S) \tag{12}$$

As $x_j$ increases from 0 towards 1, $\overline{n}_j$ increases monotonically from 0 towards $\infty$. In order that each $\overline{n}_j$ remains non-negative, we must have $x_j < 1$ for each $j$; thus either $\beta > 0$ with $\mu < r_1$ or (for finite $S$) $\beta < 0$ with $\mu > r_S$.

Another quantity of interest is $\overline{s}(n)$, the mean number of species with abundance $n$. In any given microstate ($n_1, n_2 \ldots n_S$), the number of species with population $n$ can be expressed as (Volkov et al. 2003)

$$s(n) = \sum_{j=1}^{S} I_j(n) \tag{13}$$

where the index $I_j(n)$ takes the value 1 when $n_j = n$ (with probability $p_j(n)$) and 0 otherwise (with probability $1 - p_j(n)$). Then the mean number of species with abundance $n$ is given by

$$\bar{s}(n) = \sum_{j=1}^{S}\left[1 . p_j(n) + 0.\left(1 - p_j(n)\right)\right] = \sum_{j=1}^{S} p_j(n) = \frac{1}{n+1}\sum_{j=1}^{S}\frac{x_j^{n+1}}{\left|\log_e\left(1 - x_j\right)\right|} \tag{14}$$

Neutral behaviour (n.b.) is obtained when $r_j$ and hence $x_j$ is the same for all species, so that the distribution $p_j(n) \equiv p_{\text{n.b.}}(n)$ is the same for all species. In this case each species has the same mean population size $\bar{n}_j \equiv \bar{n}_{\text{n.b.}}$ whereas the function $\bar{s}(n) = Sp_{\text{n.b.}}(n)$ varies with $n$ because it reflects statistical fluctuations in $n$ about $\bar{n}_{\text{n.b.}}$.

Eqs. (10)-(14) are the key mathematical results of our study from which follow all the results reported below.

### 4.2 *Relative frequencies of rare versus abundant species*

The distribution $\bar{s}(n)$ (Eq. 14) describes the relative frequencies of rare *vs.* abundant species. The contribution to $\bar{s}(n)$ from species $j$ has an $n$-dependence proportional to $x_j^{n}/n+1$ and $\bar{s}(n)$ is then a sum of $S$ such contributions, each with a different value of $x_j$. The total distribution $\bar{s}(n)$ is therefore characterised by many rare species and a few very common species (because each $x_j < 1$), although the precise shape of $\bar{s}(n)$ will depend on the individual values of $x_j$, *i.e.* on the per capita resource use spectrum $r_j$ (see Eq. 11b). For example, in the neutral model ($x_j = x$ for all $j$) we will have $\bar{s}(n) \propto x^n/n+1 \approx x^n/n$ for large $n$, thus reproducing the classic Fisher log series (Fisher et al., 1943).

When the values of $\bar{s}(n)$ are binned as a histogram with horizontal axis $m = \log_2 n$, the weight in bin ($m$-1,$m$) is approximately $w(n) = n\bar{s}(n)$. The contribution to $w(n)$ from species $j$ has a bell-shaped $n$-dependence given by $nx_j^{n}/(n+1)$ and $w(n)$ is a sum of $S$ such contributions, one for each species $j$. Again, the precise shape of the resulting histogram depends on the per capita resource use spectrum $r_j$. Fig. 1 illustrates the hypothetical case of a sub-linear spectrum between 0 and 1 ($\Delta = 0.0465$, $\alpha = 2/3$, $S = 101$, $\overline{N} = 1000$, $\overline{R} = 600$); the result is very similar to the empirical lognormal distribution (May 1976, Volkov et al. 2003).

### 4.3    Unimodal diversity–resource relationship at small scales

For fixed $S$ and $\overline{N}$, as $\overline{R}$ decreases towards the lower bound $\overline{R} = \overline{N}r_1$, Eq. (12) predicts that the community becomes progressively dominated by the species $j = 1$ having the lowest per capita resource use ($H_n \to 0$, Eq. 9). Conversely, when $\overline{R}$ increases towards the upper bound $\overline{R} = \overline{N}r_S$, the species $j = S$ having the highest per capita resource use increasingly dominates (again $H_n \to 0$). At the intermediate resource level

$$\overline{R}_{\text{opt}} = \frac{\overline{N}}{S} \sum_{j=1}^{S} r_j \qquad\qquad (15)$$

all $S$ species are equally abundant ($\overline{n}_j = \overline{N} / S$, $i.e.$ neutral-type behaviour) and species diversity is maximal ($H_n = \log_e S$). The predicted diversity-resource relationship is thus unimodal when $S$ and $\overline{N}$ are fixed independently of $\overline{R}$, as would be the case at small spatio-temporal scales.

From Eqs. (11b) and (12) it may be seen that the case when low-demand species dominate ($\overline{R} < \overline{R}_{\text{opt}}$) corresponds to $\beta > 0$ with $\mu < r_1$, and that when high-demand species dominate ($\overline{R} > \overline{R}_{\text{opt}}$) we have $\beta < 0$ with $\mu > r_S$. The borderline case when the mean species abundances are all equal ($\overline{R} = \overline{R}_{\text{opt}}$) corresponds to the value $\beta = 0$ (from Eq. 11b), when the $x_j$ values are the same for all species (neutral-type behaviour) even though their $r_j$ values are different.

Fig. 2 illustrates these general results for the case where $r_j$ is given by Eq. (6) with $\Delta = 0.0465$, $\alpha = 2/3$, $S = 101$ (sub-linear spectrum between 0 and 1) and $\overline{N} = 1000$, at three different resource use levels $\overline{R}$. Fig. 2(a) shows the mean relative abundances $\overline{n}_j / \overline{N}$ while Fig. 2(b) shows the corresponding unimodal $\exp(H_n) - \overline{R}$ curve. The maximum value of $\exp(H_n) = S$ occurs at the optimal resource level of $\overline{R}_{\text{opt}} = 600$ (Eq. 15). Fig. 2(b) also shows that the shape of the $\exp(H_n) - \overline{R}$ curve depends on the per capita resource use spectrum $r_j$.

When the $r_j$ levels are uniformly spaced ($\Delta = 0.01$, $\alpha = 1$), maximum diversity occurs at the lower value of $\overline{R}_{\text{opt}} = 500$.

### 4.4   Testing the unimodal pattern in a nitrogen-limited grassland community

Figure 3(a) compares the predicted relative abundances $\overline{n}_j / \overline{N}$ of the $S = 26$ species in the

CDR grasslands ($\overline{R} = 5.9$ g N m$^{-2}$ yr$^{-1}$, $\overline{N} = 62$ individuals m$^{-2}$, $r_j$ from Eq. C.1) with the relative abundance data from Harpole and Tilman (2006, figure 2a). MaxREnt predicts a realistic non-neutral species abundance distribution, with a greater relative abundance of low-N demand species.

Figure 3(b) shows the predicted change in this relative abundance distribution under increases in N use ranging from +2 to +8 g N m$^{-2}$ yr$^{-1}$ above the baseline $\overline{R} = 5.9$ g N m$^{-2}$ yr$^{-1}$.

Analogously to Fig. 2(a), MaxREnt predicts a progressive shift towards a community dominated

by species with high N demand. Figure 3(c) shows the corresponding unimodal $\exp(H_n) - \overline{R}$

curve (cf. Fig 2b); the predicted peak occurs at $\overline{R} = 9.3$ g N m$^{-2}$ yr$^{-1}$. This prediction compares well with the data of Harpole and Tilman (2006) showing a switch to higher N demand species at an increase in N addition of around +3 g N m$^{-2}$ yr$^{-1}$ ($\overline{R} \approx 9$ g N m$^{-2}$ yr$^{-1}$).

### 4.5   Monotonic diversity–resource relationship at large scales: the species-energy law

In general, when a constraint is no longer imposed, its corresponding Lagrange multiplier becomes zero (Appendix B). Thus, at larger scales we took the limit $\mu \rightarrow 0$, corresponding to a relaxation of the constraint on the total population size $\overline{N}$. The value of $\overline{N}$ was then determined

by the remaining constraint on $\overline{R}$. We also set $S = \infty$ corresponding to removal of the upper

bound on the number of possible species $S$ from which the community is assembled. Appendix D

shows that when $S = \infty$ and $\mu \to 0$, and when the metabolic scaling model (Eq. 6) is assumed, $\overline{N}$ and $\overline{R}$ are simple power law functions of the Lagrange multiplier $\beta$,

$$\overline{N} = \sum_{j=1}^{\infty} \overline{n}_j \propto \beta^{-1/\alpha} \qquad (16)$$

$$\overline{R} = \sum_{j=1}^{\infty} \overline{n}_j r_j \propto \beta^{-1/\alpha - 1} \qquad (17)$$

where in each case the power and the constant of proportionality depend on the metabolic scaling exponent $\alpha$ (see Eqs. D.1-D.3). Appendix D also shows that, in the same limit, the number of species with appreciable abundance ($S^*$), i.e. the number of species for which $\overline{n}_j \geq 1$, is proportional to $\overline{N}$ so that

$$S^* \propto \overline{N} \propto \overline{R}^{1/(1+\alpha)} \qquad (18)$$

where the last proportionality follows by eliminating $\beta$ between Eqs. (16) and (17). Again, the constant of proportionality between $S^*$ and $\overline{R}^{1/(1+\alpha)}$ depends on $\alpha$ (Eq. D.7). Fig. 4 illustrates the resulting monotonic species number–resource ($S^*-\overline{R}$) relationship.

Eq. (18) predicts a power law species number-resource relationship $S^* \propto \overline{R}^{\chi}$ with an exponent $\chi = 1/(1+\alpha)$. A power law species-energy relationship was first proposed by Wright (1983) who estimated the value $\chi = 0.62$ from the empirical relation between the number of angiosperm species and total actual evapotranspiration on 24 islands worldwide. In our derivation of Eq. (18) from MaxREnt, the value of $\chi = 1/(1+\alpha)$ depends on $\alpha$, which may be interpreted as a metabolic scaling exponent (Section 3.1). A recent general molecular model of metabolism (Demetrius, 2006) suggests that $\alpha = 2/3$ for annual plants, in which case MaxREnt predicts a species-energy exponent $\chi = 0.6$. An alternative macroscopic model of metabolism (West et al., 1997) predicts $\alpha = 3/4$, giving $\chi = 0.57$. Thus the value of $\chi$ predicted by MaxREnt is not very sensitive to the underlying metabolic model, and compares favourably with the empirical value $\chi = 0.62$ obtained by Wright (1983).

*4.6    Energetic equivalence and self-thinning behaviour*

A further simplifying approximation for the community behaviour at large scales ($S = \infty$, $\mu \to 0$) is obtained in the case of resource-rich communities, *i.e.* in the further limit $\overline{R} \to \infty$ or, equivalently, $\beta \to 0$ (see Eq. 17). In this case we can make the dual approximations $\mu \approx 0$ and $\beta r_j \ll 1$ in Eqs. (11b) and (12), which then predict that the mean population size of species $j$ is

$$\overline{n}_j = \frac{1}{\beta r_j} \frac{1}{\left| \log_e \beta r_j \right|} \approx \frac{1}{\beta r_j} \frac{1}{\left| \log_e \beta \right|} \qquad (\mu \approx 0,\ \beta r_j \ll 1) \qquad (19)$$

Hence the total rate of resource use by species $j$ is

$$\overline{n}_j r_j \approx \frac{1}{\beta} \frac{1}{\left| \log_e \beta \right|} \qquad (\mu \approx 0,\ \beta r_j \ll 1) \qquad (20)$$

which is a constant, independent of $j$. In resource-rich communities, therefore, all species (or, all sizes classes in a monoculture) contribute equally to the total community resource use $\overline{R}$ (Eq. 5). This behaviour, known as energetic equivalence (Allen et al. 2002; Enquist et al., 1998), implies that the total resource use by each species is the same.

Combining energetic equivalence (Eq. 20) and metabolic scaling (Eq. 7) gives an inverse power law relationship (self-thinning law) between species mass and population size,

$$m_j \propto \overline{n}_j^{\,-1/\alpha} \qquad (21)$$

for which the values $\alpha = 2/3$ (Demetrius, 2006) and $\alpha = 3/4$ (West *et al.* 1997) give, respectively, –3/2 and –4/3 self-thinning power laws.

Our derivation of energetic equivalence from MaxREnt implies that it is the most probable distribution of resource use among species in resource-rich communities. But MaxREnt also predicts that energetic equivalence (and hence also Eq. 21) will break down in resource-poor communities, when $\overline{R}$ is not large enough for the approximation in Eq. (20) to remain valid.

## 5.    Discussion

### 5.1    *Reconciling contrasting ecological patterns*

Our aim has been to show how statistical mechanics explains and reconciles a disparate collection of empirical ecological patterns. The common principle linking these patterns is that community-level behaviour represents the most probable way in which a large number of internal degrees of freedom (individuals) will self-organise when subjected to a relatively small number of environmental constraints (space and resources). By encoding only those constraints into the microstate distribution $p$, MaxREnt predicts the reproducible (= most probable) community-level behaviour under those constraints.

In this way MaxREnt explains and reconciles the contrasting relationships between diversity and resource level at different scales (Figs. 2-4). Recalling the quote from Gaston (2000) in Section 1, the 'reconciliation of the patterns in biodiversity that are observed at different scales', as provided by MaxREnt, leads to the following 'insights into their determinants': diversity–resource patterns are manifestations of the most probable way in which communities could assemble under given environmental constraints, the contrasting patterns (unimodal *vs.* monotonic) reflecting the different constraints ($S$ finite and $\overline{N}$ fixed *vs.* $S = \infty$ and $\overline{N}$ free) that operate on different scales.

Previous explanations of the positive correlation of taxonomic richness with climate at large scales have focused on particular mechanisms, involving, for example, the effects of climate on net primary productivity and population size, on physiological tolerance, or on speciation rates (e.g. Currie et al., 2004). In contrast, MaxREnt provides a statistical explanation for this correlation based on the most likely allocation of community resource use among different species. In general, the species number–resource ($S$*–$\overline{R}$) relationship predicted by MaxEnt at large scales is monotonic because the species resource-use spectrum $r_j$ is unbounded ($S = \infty$) and therefore, as $\overline{R}$ increases, the most likely species abundance distribution $\overline{n}_j$, while always having greatest weight at $j = 0$, spreads upwards to occupy higher levels of the spectrum.

In the particular case where $r_j$ is related allometrically to body size (metabolic scaling, Eq. 7), the resulting $S^* - \overline{R}$ relationship reproduces the species-energy power law (Wright, 1983), in which the empirical exponent $\chi \approx 0.6$ can be explained in terms of the metabolic scaling exponent $\alpha \approx 2/3$ through the prediction $\chi = 1/(1 + \alpha)$.

MaxREnt also reconciles neutral and non-neutral behaviours. Neutral behaviour obviously applies when all species have the same per capita resource use $r_j = r$, when all species have the same mean abundances $\overline{n}_j$. By introducing species differences in the $r_j$ values, we are able to reproduce realistic non-neutral abundance distributions (Harpole and Tilman, 2006), as illustrated in Fig. 3. However, even when the $r_j$ values are different, the predicted unimodal diversity-resource pattern for finite $S$ (Figs. 2 and 3) implies neutral-type behaviour at or near the peak in diversity, when the value of $\overline{R}$ is optimal (Eq. 15).

Part of the difficulty in distinguishing between different mechanisms of community assembly is that they predict very similar patterns for the relative frequency of rare *vs.* abundant species (e.g. Chave, 2004; Chave et al., 2002; Volkov et al., 2003). MaxREnt likewise predicts a pattern close to the empirical lognormal distribution (Fig. 1), although its precise form depends on the particular resource use spectrum $r_j$ that is used. Thus, the relative species abundance distribution alone does not strongly discriminate between MaxREnt and other models. Rather, it is the totality of the ecological patterns unified by MaxREnt that provides evidence in its favour.

Moreover, MaxREnt provides a statistical perspective on community assembly which contrasts with previous dynamical models. MaxREnt explains community assembly in terms of the most probable species abundance distribution that could occur under given environmental constraints, without assuming anything about the internal population dynamics. The rationale here is that the rates of birth, death and dispersal are themselves stochastic variables, whose most probable values are largely conditioned by the environmental constraints, so that ultimately only the latter need to be encoded into the microstate distribution $p(n_1, n_2 \ldots n_S)$.

In addition to reconciling contrasting diversity-resource patterns and neutral *vs.* non-neutral behaviours, MaxREnt leads to a statistical interpretation of energetic equivalence (e.g.

Allen et al., 2002; Damuth, 1987; Enquist et al., 1998) and the self-thinning power law (e.g. Dewar, 1993, 1999; Enquist et al., 1998). In previous studies (Dewar, 1993; Enquist et al., 1998), energetic equivalence was assumed as a starting hypothesis and then combined with metabolic scaling to obtain a self-thinning law between body size (or mass) and population density. Here, using MaxREnt, we have gone one step further back in this chain of reasoning, by deriving energetic equivalence as the most probable distribution of resource use among species in a resource-rich community. Note that energetic equivalence and metabolic scaling are logically independent. Within the approximations used to derive Eq. (20), energetic equivalence is a very general statistical result, valid for arbitrary fixed values of $r_j$. In contrast, metabolic scaling (Eq. 7) is a particular model for how $r_j$ relates to body size, and only for that model of $r_j$ does energetic equivalence then imply the self-thinning power law (Eq. 21).

Finally, what insight does statistical mechanics provide into why there are so many species on Earth (e.g. Tilman, 2000)? Statistical mechanics tells us that, all else being equal, probability distributions that are more spread-out (diverse) can be realized in many more ways than narrowly-peaked distributions (e.g. Jaynes, 2003). Hence maximally-diverse species abundance distributions are just what we expect to observe simply because they are the most likely ones. MaxREnt is therefore a useful conceptual and practical step forward for ecology, because it suggests a statistical explanation for the great diversity of species on Earth as well as providing a quantitative prediction of diversity under given environmental constraints.

*5.2 Relation of MaxREnt to some previous applications of statistical mechanics to ecology*

Li et al. (2000) derived a mass–density (self-thinning) relationship by applying MaxEnt to construct the distribution of nearest-neighbour distances between individual plants in a population, subject to constraints on the average (population-level) nearest-neighbour area and the average 'interaction intensity' (an index of spatial competition between individuals). In contrast, we make no explicit assumption about spatial competition. While spatial interactions do exist at the individual level, we have shown that the community-level resource constraint $\overline{R}$ (Eq. 5), when combined with metabolic scaling (Eq. 7), suffices to explain self-thinning patterns, suggesting that the details of those spatial interactions do not influence patterns at the community level.

Shipley et al. (2006) used MaxEnt to infer the relative distribution of above-ground dry biomass among 30 species in herbaceous communities, subject to the mean (community-aggregated) values of 8 functional traits (which included seed maturation date, specific leaf area, above-ground biomass and height) at 12 sites in a chronosequence. Predicted and observed above-ground biomass distributions were similar. Our approach differs from Shipley et al. (2006), as it does from Li et al. (2000), in that $\overline{N}$ and $\overline{R}$ give a more direct representation of the actual environmental constraints (space and resources) to which communities are subjected. Also, the main objective of our study has been to explain and unify the ecological patterns observed in multi-species communities generally, rather than to explain a particular dataset.

MaxREnt also differs crucially from both of these applications of MaxEnt in the inclusion of the prior distribution $q$. In our application of MaxREnt to $p(\boldsymbol{n})$, the ignorance prior $q(\boldsymbol{n})$ leads to the pre-factor $1/(n+1)$ in Eq. (11a), and hence to a Fisher log series behaviour for $\overline{s}(n)$ at large $n$. Some previous statistical mechanical models of diversity – although not explicitly MaxREnt – have effectively incorporated a prior distribution $q(\boldsymbol{n})$ into $p(\boldsymbol{n})$ through specific assumptions about the underlying population dynamics (Volkov et al., 2004, 2005). For example, Volkov et al. (2004) obtained $q(\boldsymbol{n}) = \prod_j 1/n_j$ by assuming constant per capita birth and death rates for all species. In a modification to this model, Volkov et al. (2005) obtained $q(\boldsymbol{n}) = \prod_j \left(n_j + c\right)^{-1}$ in which the parameter $c$ describes density-dependent corrections. In contrast, by construction our ignorance prior $q(\boldsymbol{n}) = \prod_j \left(n_j + 1\right)^{-1}$ makes no assumptions whatsoever about the underlying population dynamics. For the same reason our study also differs from the application of statistical mechanics to population dynamics developed by Demetrius (1983, 1997).

Mathematically, our prior is equivalent to Volkov et al.'s (2005) density-dependent model with $c = 1$, which can therefore be viewed as a null hypothesis for population dynamics. Moreover, the null hypothesis $c = 1$ lies well within the range of values of $c$ fitted to six tropical forest data sets (Table 1 of Volkov et al., 2005), which suggests that information about the underlying population dynamics is (at least for those data sets) largely irrelevant to prediction of species abundance patterns. In systems for which density-dependent effects are manifested at the community level, we suggest they might be predicted by modifying the simple resource–use model (Eq. 5) rather than the ignorance prior $q(\boldsymbol{n})$. For example, intra- and inter-species interactions among individuals can be incorporated by making each $r_j$ a function of all the $n_k$ ($k = 1, 2 \dots S$).

### 5.3    The generic origin of ecological patterns

The types of statistical distribution found in studies of ecological communities (lognormal abundance distributions, species-energy power laws, etc.) are not unique to ecology but are seen also in, for example, economies, physical systems and social systems (May, 1975; Nekola and Brown, 2007; Preston, 1950). Therefore the explanation of these patterns cannot be unique to ecology either. We suggest that MaxREnt provides such a generic explanation.

Indeed, many features of our model and its statistical behaviour have mathematical similarities with the statistical behaviour of thermodynamic systems (e.g. Reichl, 1980). As Volkov et al. (2004) noted, individuals in a community are analogous to quantum mechanical particles called bosons: both represent system components whose number $n_j$ ranges in principle from zero to infinity (for fermions $n_j = 0$ or 1 only). The distribution of individuals among species with different per capita resource use $r_j$ is analogous to the distribution of bosons among different energy levels. The constraints on $\overline{N}$ and $\overline{R}$ (Eqs. 4 and 5) are then mathematically identical to the constraints on the particle number ($\overline{N}$) and total internal energy ($\overline{E}$) of a system of bosons in thermal equilibrium with a heat reservoir.

As a result, the statistical behaviour predicted by MaxREnt under those constraints shows mathematical similarities with the statistical behaviour of bosons (Reichl, 1980): (i) our Lagrange multiplier parameters β and μ bear the same mathematical relation to $\overline{R}$ and $\overline{N}$ as the inverse temperature and chemical potential do to $\overline{E}$ and $\overline{N}$ in thermodynamics (Appendix E);

(ii) the case when high-demand species have greater mean abundance than low-demand species ($\overline{R} > \overline{R}_{\text{opt}}$, $\beta < 0$, $\mu > r_S$, Fig. 3b) is analogous to a negative-temperature thermodynamic system in which high energy levels are more populated than low energy levels; (iii) energetic equivalence (Eq. 20), which is valid for large $\overline{R}$, is analogous to the principle of equipartition of energy valid for a thermodynamic system at high temperature (the 'classical limit'), when $\overline{E}$ is distributed equally among the energy levels; (iv) just as equipartition of energy breaks down at low temperatures, so energetic equivalence breaks down when $\overline{R}$ is small.

Of course individual organisms are not bosons. The point we wish to make is that the statistical behaviour of a collection of individual organisms is mathematically similar to the statistical behaviour of a collection of bosons, because both systems are composed of an integer number ($n_j$) of entities with different properties ($r_j$) and subjected to environmental constraints of the same mathematical form (Eqs. 4 and 5). The analogy is fruitful because it shows how ecological patterns can be understood in terms of the generic statistical behaviour of large assemblages of individual entities, independently of the precise nature of those entities.

6. **Conclusion**

In addition to its rôle as a tool for statistical inference, MaxREnt also provides a statistical mechanical tool for predicting the most probable behaviour of complex systems with many degrees of freedom under given constraints. We have shown how MaxREnt explains and unifies a disparate collection of ecological patterns in multi-species communities. The explanation of those patterns is not unique to ecology but rather reflects the generic statistical behaviour of complex systems with many degrees of freedom under very general types of environmental constraints.

**Appendix A: the prior distribution $q(\boldsymbol{n})$, Eq. (8)**

We seek the prior distribution $q\left(n_1, n_2 \ldots n_S\right)$ that describes complete uncertainty about the species populations $n_1$, $n_2$ *etc*. For each species $j$, the population $n_j$ can take any one of the infinite set of values 0, 1, 2 …. $\infty$ and so $m_j \equiv n_j + 1$ can take any one of the values 1, 2 …. $\infty$. Rissanen (1983) has derived a universal ignorance prior for the positive integers $m > 0$ based on considerations of the minimum length of binary strings required to code $m$. The result has the form $q(m) \propto 2^{-\log^* m}$ where $\log^* m \equiv \log_2 m + \log_2 \log_2 m + \ldots$. in which the sum involves only non-negative terms, whose number is therefore finite. The distribution $q(m)$ is summable *i.e.* $\sum\limits_{m=1}^{\infty} q(m)$ is finite.

A useful approximation to $\log^* m$ (sufficient for our purposes) is given by its first term, leading to the Jeffreys prior $q(m) \propto 1/m$. Because complete uncertainty of the numbers $n_1$, $n_2$ *etc*. excludes any informative correlations between them, we then have the ignorance prior

$$q(\boldsymbol{n}) = q\left(n_1, n_2 \ldots n_S\right) = \prod\limits_{j=1}^{S} \frac{1}{n_j + 1} \qquad (A.1)$$

which is Eq. (8) of the main text. Although this approximate expression for $q(\boldsymbol{n})$ is not summable, within the context of MaxREnt subject to constraints it leads to a posterior distribution $p(\boldsymbol{n})$ that is summable.

**Appendix B: general solution of MaxREnt for $p(\boldsymbol{n})$**

The maximisation of $H\left(p \| q\right)$ with respect to $p(\boldsymbol{n})$ subject to normalisation of $p$ and to Eqs. (4) and (5) is accomplished by constructing the objective function

$$\Psi \equiv -\sum\limits_{\boldsymbol{n}} p(\boldsymbol{n}) \log_e \frac{p(\boldsymbol{n})}{q(\boldsymbol{n})}$$

$$- \lambda_0 \left(1 - \sum\limits_{\boldsymbol{n}} p(\boldsymbol{n})\right) - \lambda_1 \left(\overline{N} - \sum\limits_{\boldsymbol{n}} p(\boldsymbol{n}) \sum\limits_{j=1}^{S} n_j\right) - \lambda_2 \left(\overline{R} - \sum\limits_{\boldsymbol{n}} p(\boldsymbol{n}) \sum\limits_{j=1}^{S} n_j r_j\right) \qquad (B.1)$$

where $\lambda_0$, $\lambda_1$ and $\lambda_2$ are the respective Lagrange multipliers associated with the constraints. Here the sum on $\boldsymbol{n} = (n_1, n_2 \ldots n_S)$ denotes a sum over each $n_j = 0, 1, 2 \ldots \infty$ independently. Setting $\partial \Psi / \partial p(\boldsymbol{n}) = 0$ gives the solution

$$p(\boldsymbol{n}) = q(\boldsymbol{n})\exp\left[-1 + \lambda_0 + \lambda_1 \sum_{j=1}^{S} n_j + \lambda_2 \sum_{j=1}^{S} n_j r_j\right] = \frac{1}{Z}\prod_{j=1}^{S}\frac{e^{(\lambda_1 + \lambda_2 r_j)n_j}}{n_j + 1} \tag{B.2}$$

where $Z = \exp(1 - \lambda_0)$ and we have substituted $q(\boldsymbol{n})$ from Eq. (A.1). The normalisation constraint implies that

$$Z = \sum_{n_1, n_2 \ldots n_S = 0}^{\infty}\left(\prod_{j=1}^{S}\frac{e^{(\lambda_1 + \lambda_2 r_j)n_j}}{n_j + 1}\right) = \prod_{j=1}^{S}\left(\sum_{n=0}^{\infty}\frac{e^{(\lambda_1 + \lambda_2 r_j)n}}{n + 1}\right) = \prod_{j=1}^{S}\left(\sum_{n=0}^{\infty}\frac{x_j^{\,n}}{n + 1}\right) = \prod_{j=1}^{S} Z_j \tag{B.3}$$

where we have introduced the notation

$$x_j = e^{\lambda_1 + \lambda_2 r_j} \tag{B.4}$$

$$Z_j = \sum_{n=0}^{\infty}\frac{x_j^{\,n}}{n + 1} = \frac{-\log_e(1 - x_j)}{x_j} = \frac{\left|\log_e(1 - x_j)\right|}{x_j} \tag{B.5}$$

The last equality in Eq. (B.5) follows from the fact that $x_j < 1$ in order that the sum over $n$ converges. From Eqs. (B.2)-(B.4) it follows that $p(\boldsymbol{n})$ has the form (cf. Eq. 10)

$$p(\boldsymbol{n}) = \prod_{j=1}^{S} p_j(n_j) \tag{B.6}$$

where (cf. Eq. 11a)

$$p_j(n) = \frac{1}{Z_j}\frac{x_j^{\,n}}{n + 1} = \frac{x_j^{\,n+1}}{n + 1}\frac{1}{\left|\log_e(1 - x_j)\right|} \tag{B.7}$$

In order to highlight the mathematical similarities with thermodynamic systems (Appendix E), it is useful to reparameterise the Lagrange multipliers as $\lambda_1 \equiv \beta\mu$ and $\lambda_2 \equiv -\beta$ so that Eq. (B.4) becomes (cf. Eq. 11b)

$$x_j \equiv e^{-\beta(r_j - \mu)} \tag{B.8}$$

The parameters $\beta$ and $\mu$ are mathematically analogous to the inverse temperature and chemical potential, respectively, in equilibrium thermodynamics (Appendix E). When the constraint on $\overline{N}$ (Eq. 4) is relaxed, the corresponding Lagrange multiplier $\lambda_1 \equiv \beta\mu$ is set to zero. This corresponds to taking the limit $\mu \to 0$.

**Appendix C: parameterising the fixed-S model for a nitrogen-limited grassland**

We used published data from nitrogen-limited grasslands of the Cedar Creek Long Term Ecological Research site, Minnesota (CDR E014 oldfields 32, 35 and 72) (Harpole and Tilman, 2006; Tilman, 1984). We considered the community-level constraints to be the mean number of individuals ($\overline{N}$) and the annual rate of soil N use ($\overline{R}$).

$\overline{N}$ was given as 62 individuals per m$^2$ (Harpole and Tilman, 2006). We estimated $\overline{R}$ (g N m$^{-2}$ yr$^{-1}$) from mineralised soil N concentration ([N]$_{min}$, mg N kg$^{-1}$ soil), soil bulk density ($\rho$, g cm$^{-3}$) and soil depth ($D$, m), by assuming that all mineralised soil N was extracted by the community in one year, so that $\overline{R} = [N]_{min} \rho D$. [N]$_{min}$ and $\rho$ were estimated using empirical relationships to total soil N concentration ([N]$_{tot}$, mg N kg$^{-1}$ soil) established on other CDR oldfields with similar species composition: respectively, [N]$_{min}$ = 0.029[N]$_{tot}$ + 17.69 (Pastor et al., 1987) and $\rho$ = 1.509 − 0.000103[N]$_{tot}$ (Wedin and Tilman, 1990). [N]$_{tot}$ in the three oldfields was calculated for 2002 using an empirical relationship to field age ($a$, yr) : [N]$_{tot}$ = 6.84$a$ + 369.19 (Inouye et al., 1987). $D$ was 0.1 m (Inouye et al., 1987). We thus obtained the estimate $\overline{R}$ = 5.9 g N m$^{-2}$ yr$^{-1}$, corresponding to a low level of N availability for grasslands (Oomes et al., 1997; Warren and Whitehead, 1988; Woodmansee and Duncan, 1980).

The oldfield community is composed of a fixed spectrum of $S$ = 26 potential species (Harpole and Tilman, 2006), each characterised by a species-specific per capita N use, $r_j$ (g N yr$^{-1}$ individual$^{-1}$). We assumed that $r_j$ was proportional to $R_j$* (mg N kg$^{-1}$ soil), an index of the competitiveness of species $j$ for N (Harpole and Tilman, 2006, table 1), measured as the amount of available N remaining in the soil once a monoculture has reached equilibrium. $R_j$* is considered to be the requirement of N for survival, beyond which an increase in N availability results in an increase in species abundance within the community (Tilman, 1990). $R_j$* values were obtained for the major species of the CDR grassland community (Craine et al., 2002;

Harpole and Tilman, 2006). Studies of N uptake in natural or mesocosm grasslands (Chaneton et al., 1996; Oomes et al., 1997; Woodmansee and Duncan, 1980) were used to estimate $r_j$ and its ratio to $R_j*$ values. We thus derived the relationship

$$r_j = f R_j^* \rho D / \overline{N} \qquad (C.1)$$

in which $\overline{N} = 62$ individuals per m$^2$ is the measured community density, and $f = 230$ converts N use in survival conditions to that in growing conditions. We used Eq. (C.1) to estimate each $r_j$ ($j = 1, 2, \dots 26$) from the $R_j*$ values given by Harpole and Tilman (2006, table 1).

**Appendix D: community behaviour at large scales ($S = \infty$, $\mu \rightarrow 0$)**

For $S = \infty$ and $\alpha < 1$, we can approximate the infinite sums over $j$ in Eqs. (4) and (5) by integrals over the per capita resource use, $r$. Let $\rho(r)dr$ be the number of species for which $r_j$ lies in the range $(r, r+dr)$. The sums over $j$ can then be approximated by integrals over $r$ with weight function $\rho(r)$. By definition $1/\rho(r) = dr/dj$ so that Eq. (6) gives $\rho(r) = r^{1/\alpha-1}/(\alpha\Delta^{1/\alpha})$. By introducing the change of variables $\lambda \equiv \exp(\beta\mu)$ and $y = \beta r$, then substituting Eq. (12) for $\overline{n}_j$ into Eqs. (4) and (5), and setting $\lambda \equiv \exp(\beta\mu) \approx 1$ in the limit $\mu \rightarrow 0$, we obtain the following approximations:

$$\overline{N} = \sum_{j=1}^{\infty} \overline{n}_j \approx \beta^{-1/\alpha} I_0(\alpha) \qquad (D.1)$$

$$\overline{R} = \sum_{j=1}^{\infty} \overline{n}_j r_j \approx \beta^{-1/\alpha-1} I_1(\alpha) \qquad (D.2)$$

where $I_0$ and $I_1$ are given as functions of $\alpha$ by the integral expression

$$I_k(\alpha) = \frac{1}{\alpha\Delta^{1/\alpha}} \int_0^{\infty} dy\, y^{1/\alpha+k-1} f\left(e^{-y}\right) \qquad (k = 0 \text{ or } 1) \qquad (D.3)$$

in which

$$f(x) = \frac{x}{1-x} \frac{1}{\left|\log_e(1-x)\right|} - 1 \qquad (D.4)$$

The number of species with appreciable abundance, $S*$, can be shown to be proportional to $\overline{N}$ as follows. From Eq. (12), the mean abundances $\overline{n}_j$ decrease with increasing species index $j$ ($\beta > 0$ in the case $S = \infty$ we are considering here). By definition, $S^*$ is the value of the species index $j$ at which $\bar{n}_j$ falls to the value one. From Eq. (12) the condition $\bar{n}_{j=S^*} = 1$ may be solved numerically to give $x_{j=S^*} = 0.716$, and from Eq. (11b) with $\mu \approx 0$, this corresponds to the condition $\beta r_{j=S^*} = 0.335$. Substituting $r_{j=S^*} = \Delta(S*-1)^\alpha \approx \Delta(S*)^\alpha$ from Eq. (6), and re-arranging, we then find

$$S* \approx \left( \frac{0.335}{\beta \Delta} \right)^{1/\alpha} \tag{D.5}$$

which, we see from Eq. (D.1), is proportional to $\bar{N}$. By eliminating $\beta$ between Eqs. (D.2) and (D.5), we obtain the species-energy power law

$$S* \approx b(\alpha) \bar{R}^{1/(1+\alpha)} \tag{D.6}$$

where the constant of proportionality $b(\alpha)$ depends on the metabolic scaling exponent $\alpha$:

$$b(\alpha) = \left( \frac{0.335}{\Delta} \right)^{1/\alpha} I_1(\alpha)^{-1/(1+\alpha)} \tag{D.7}$$

**Appendix E: ecological interpretation of the Lagrange multiplier parameters μ and β**

Mathematically, the parameters $\mu$ and $\beta$ are simply auxilliary variables, each of which is determined by the ecological constraints on $\bar{N}$ and $\bar{R}$ (Eqs. 4 and 5). Nevertheless it is useful to interpret $\mu$ and $\beta$ ecologically by analogy with the role the corresponding parameters play in thermodynamic systems constrained by total particle number $\bar{N}$ and internal energy $\bar{E}$ (e.g. Reichl, 1980).

By substituting the MaxREnt solution for $p(\boldsymbol{n})$ from Eq. (B.2) into Eq. (2), the maximised value of the relative entropy $H(p\|q)$ becomes a function of the values of $\bar{N}$ and $\bar{R}$ under which it was maximised:

$$\hat{H}(\bar{N}, \bar{R}) \equiv \max\left( -\sum_{\boldsymbol{n}} p(\boldsymbol{n}) \log_e \frac{p(\boldsymbol{n})}{q(\boldsymbol{n})} \right)\Bigg|_{\bar{N}, \bar{R}} = \log_e Z + \beta(\bar{R} - \mu\bar{N}) \tag{E.1}$$

where $Z$ is given by Eq. (B.3). By taking derivatives of Eq. (E.1) one may show that

$$\beta = \frac{\partial \hat{H}(\overline{N}, \overline{R})}{\partial \overline{R}} \tag{E.2}$$

and

$$\mu = -\frac{1}{\beta} \frac{\partial \hat{H}(\overline{N}, \overline{R})}{\partial \overline{N}} \tag{E.3}$$

These relationships – which hold quite generally irrespective of the resource-use model (Eq. 5) – are mathematically identical to their thermodynamic counterparts (Reichl, 1980, Chapt. 2) $1/T = \partial \hat{S}/\partial \overline{E}$ and $\mu'/T = -\partial \hat{S}/\partial \overline{N}$ (where $T$ = temperature, $\hat{S}$ = entropy , $\overline{E}$ = internal energy, $\mu'$ = chemical potential) through the correspondences $\hat{H} \leftrightarrow \hat{S}/k_B$ ($k_B$ = Boltzmann's constant), $\overline{R} \leftrightarrow \overline{E}$ , $\beta \leftrightarrow 1/k_B T$ and $\mu \leftrightarrow \mu'$ . That $\beta$ is the analogue of inverse temperature ($1/k_B T$) can also be seen directly from Eqs. (16) and (17) which imply that $1/\beta$ is proportional to the average resource use per individual ( $\overline{R}/\overline{N}$ ), analogous to the result that temperature is proportional to the average energy per degree of freedom ( $\overline{E}/\overline{N}$ ) in thermodynamic systems.

Another insight into the rôle of $\mu$ and $\beta$ as ecological analogues of chemical potential and inverse temperature ($1/k_B T$) is revealed by considering two initially isolated communities (labelled 1 and 2) with arbitrary initial population sizes and resource use rates $\left(\overline{N}_{1i}, \overline{R}_{1i}\right)$ and $\left(\overline{N}_{2i}, \overline{R}_{2i}\right)$, respectively. Suppose they are then allowed to interact by exchanging individuals and resources in such a way that the combined population size $\overline{N}_{1i} + \overline{N}_{2i} = \overline{N}$ and resource use rate $\overline{R}_{1i} + \overline{R}_{2i} = \overline{R}$ remain constant. What are the final population sizes and resource use rates, $\left(\overline{N}_{1f}, \overline{R}_{1f}\right)$ and $\left(\overline{N} - \overline{N}_{1f}, \overline{R} - \overline{R}_{1f}\right)$, of each community?

If the interaction between the two systems is sufficiently weak, the microstate probability distribution $p$ for the combined community is the product of those for each community, $p = p_1 p_2$. Similarly $q = q_1 q_2$. In that case Eq. (2) implies that the relative entropy of the combined system is just the sum of the relative entropies of each system, and as a function of $\overline{N}_{1f}$ and $\overline{R}_{1f}$ this result may be written in the form

$$\hat{H}_{\text{total}}(\overline{N}_{1f}, \overline{R}_{1f}) = \hat{H}_1(\overline{N}_{1f}, \overline{R}_{1f}) + \hat{H}_2(\overline{N} - \overline{N}_{1f}, \overline{R} - \overline{R}_{1f}) \tag{E.4}$$

By maximising $\hat{H}_{\text{total}}(\overline{N}_{1f}, \overline{R}_{1f})$ with respect to $\overline{N}_{1f}$ and $\overline{R}_{1f}$ to find the most probable final state, one finds that the following relations between partial derivatives must hold at $(\overline{N}_{1f}, \overline{R}_{1f})$:

$$\frac{\partial \hat{H}_1(\overline{N}_1, \overline{R}_1)}{\partial \overline{R}_1} = \frac{\partial \hat{H}_2(\overline{N}_2, \overline{R}_2)}{\partial \overline{R}_2} \tag{E.5}$$

and

$$\frac{\partial \hat{H}_1(\overline{N}_1, \overline{R}_1)}{\partial \overline{N}_1} = \frac{\partial \hat{H}_2(\overline{N}_2, \overline{R}_2)}{\partial \overline{N}_2} \tag{E.6}$$

By comparing these relations with Eqs. (E.2) and (E.3) we see that the most probable final state is the one for which $\beta_1 = \beta_2$ and $\mu_1 = \mu_2$. Therefore, as the combined system moves from a less

probable initial state to the most probable final state, any initial difference in $\beta$ between the two

communities will tend to zero through the reallocation of resources between them, while any

initial difference in $\mu$ will tend to zero through an exchange of individuals. Thus differences in $\beta$

and $\mu$ are measures of the tendency for resources and individuals to be reallocated among interacting communities, and as such they play rôles analogous to the temperature and chemical potential, respectively, of a thermodynamic system with constant internal energy $\overline{E}$ and particle number $\overline{N}$.

### Acknowledgements

R.D. thanks Jayanth Banavar, John Harte, Amos Maritan, Alan McKane and Igor Volkov

for stimulating discussions during the Workshop on Quantitative Ecology, Abdus-Salam

International Centre for Theoretical Physics, 9-20 May, 2005. R.D. also thanks Amos Maritan

for hosting a visit to Padua University (23-25 January, 2007) during which he indicated the role of relative entropy maximisation in explaining species abundance patterns (Banavar and Maritan,

http://xxx.lanl.gov/abs/cond-mat/0703622). Lloyd Demetrius, Amos Maritan, Cathy Neill, Bill

Shipley and an anonymous reviewer provided many valuable comments on the manuscript.

**Figure Legends**

**Fig. 1.** Relative frequency of rare *vs.* abundant species. Histogram: a plot of $w(n) = n\bar{s}(n)$, where $\bar{s}(n)$ is the mean number of species with population $n$, on a $\log_2 n$ scale, for a community of $\bar{N} = 1000$ individuals with resource use $\bar{R} = 600$. $\bar{s}(n)$ was calculated from Eqs. (11b) and (14) in which $r_j$ is given by Eq. (6) with $\Delta = 0.0465$, $\alpha = 2/3$, $S = 101$ (sub-linear spectrum between 0 and 1) and the Lagrange multiplier parameters $\mu$ and $\beta$ are adjusted to satisfy the constraint Eqs. (4) and (5). With this parameterisation all $S$ species are equally abundant (neutral pattern, see Fig. 2a). Curve: a lognormal fit of the form

$$w(n) = W_0 \exp\left\{-\left(\log_2 n - \log_2 n_0\right)^2 / 2\sigma^2\right\}$$ with $W_0 = 20.4$, $\log_2 n_0 = 2.4$, $\sigma = 2.3$.

**Fig. 2.** Predictions for fixed $S$ and $\bar{N}$. (a) Percentage relative abundances $\bar{n}_j / \bar{N} \times 100\%$ ($j = 1 \ldots S$) of $S = 101$ species in a community of $\bar{N} = 1000$ individuals at three community resource use levels ($\bar{R} = 500$, 600 and 700). For each value of $\bar{R}$, $\bar{n}_j$ was calculated from Eq. (12); other details as in Fig. 1. The case $\bar{R} = 600$ gives a neutral-type species abundance distribution ($\bar{n}_j$ independent of $j$) even though species have different per capita resource use $r_j$; the other two cases give non-neutral distributions. (b) Solid curve: the corresponding unimodal relationship between the exponential Shannon diversity index, $\exp(H_n)$ (Eq. 9), and the community resource use $\bar{R}$. Maximum diversity ($\exp(H_n) = S$) occurs at $\bar{R}_{\text{opt}} = 600$ (Eq. 15) when all species have equal mean abundances (neutral-type behaviour). Broken curve: corresponding prediction for $\Delta = 0.01$, $\alpha = 1.0$ (linear spectrum between 0 and 1) for which maximum diversity occurs at $\bar{R}_{\text{opt}} = 500$.

**Fig. 3.** Testing the predictions for fixed $S$ and $\overline{N}$. (a) Predicted and measured relative species abundances as a function of the species competitiveness index $R_j$*. Solid line: MaxREnt prediction ($S = 26$, $\overline{R} = 5.9$ g N m$^{-2}$ yr$^{-1}$, $\overline{N} = 62$ individuals m$^{-2}$, $r_j$ from Eq. C.1). Boxes: data from Harpole and Tilman (2006). (b) Effect on relative species abundances of an increase in community N use (+2 to +8 g N m$^{-2}$ yr$^{-1}$) above the baseline rate $\overline{R} = 5.9$ g N m$^{-2}$ yr$^{-1}$ with $S$ and $\overline{N}$ fixed. (c) Effect of community N use on the exponential Shannon diversity index, exp($H_n$). The modelled peak in diversity at $\overline{R}_{\text{opt}}^{\text{(mod.)}} = 9.3$ g N m$^{-2}$ yr$^{-1}$ lies close to the value $\overline{R}_{\text{opt}}^{\text{(obs.)}} \approx$ 9 g N m$^{-2}$ yr$^{-1}$ at which a switch to communities dominated by high-N demand species was observed (Harpole and Tilman, 2006).

**Fig. 4.** Behaviour at large spatio-temporal scales. The monotonic species number-resource relationship ($S$*$-\overline{R}$) predicted by Eqs. (D.6) and (D.7) for three values of $\alpha$ and $\Delta = 1$. $S$* is the number of species whose mean population is greater than or equal to one ($\overline{n}_j \geq 1$). The predicted power-law relationship $S* \propto \overline{R}^{1/(1+\alpha)}$ has an exponent that is relatively insensitive to the metabolic scaling exponent $\alpha$.

Fig. 1

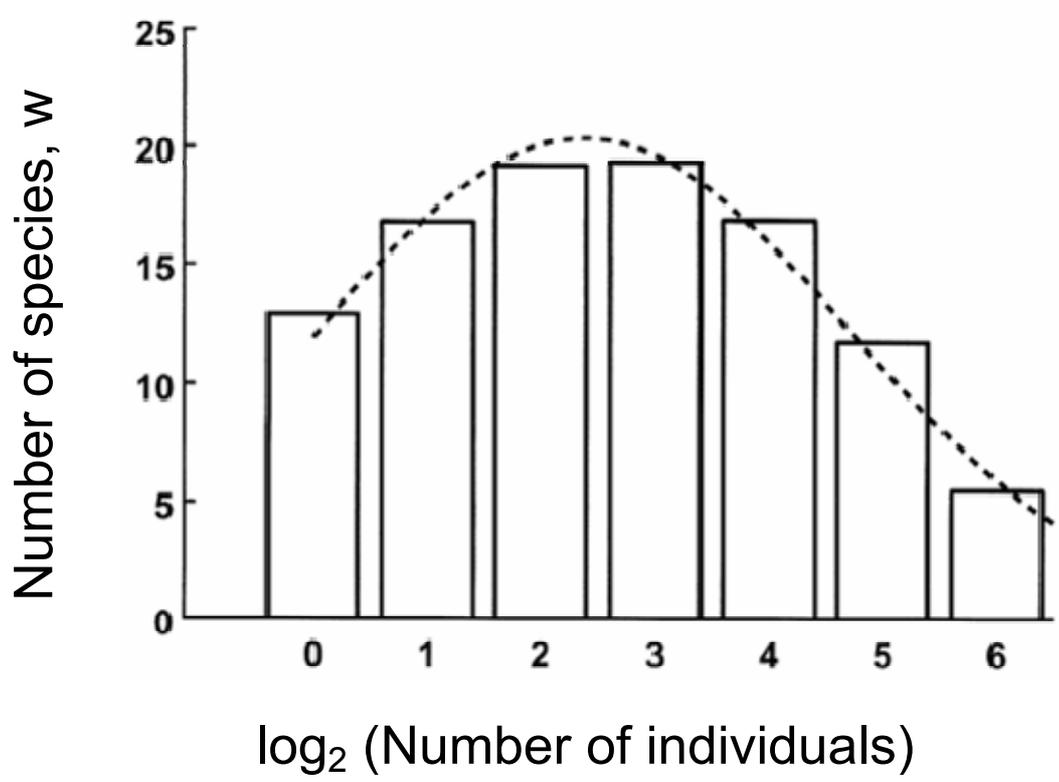

(a)

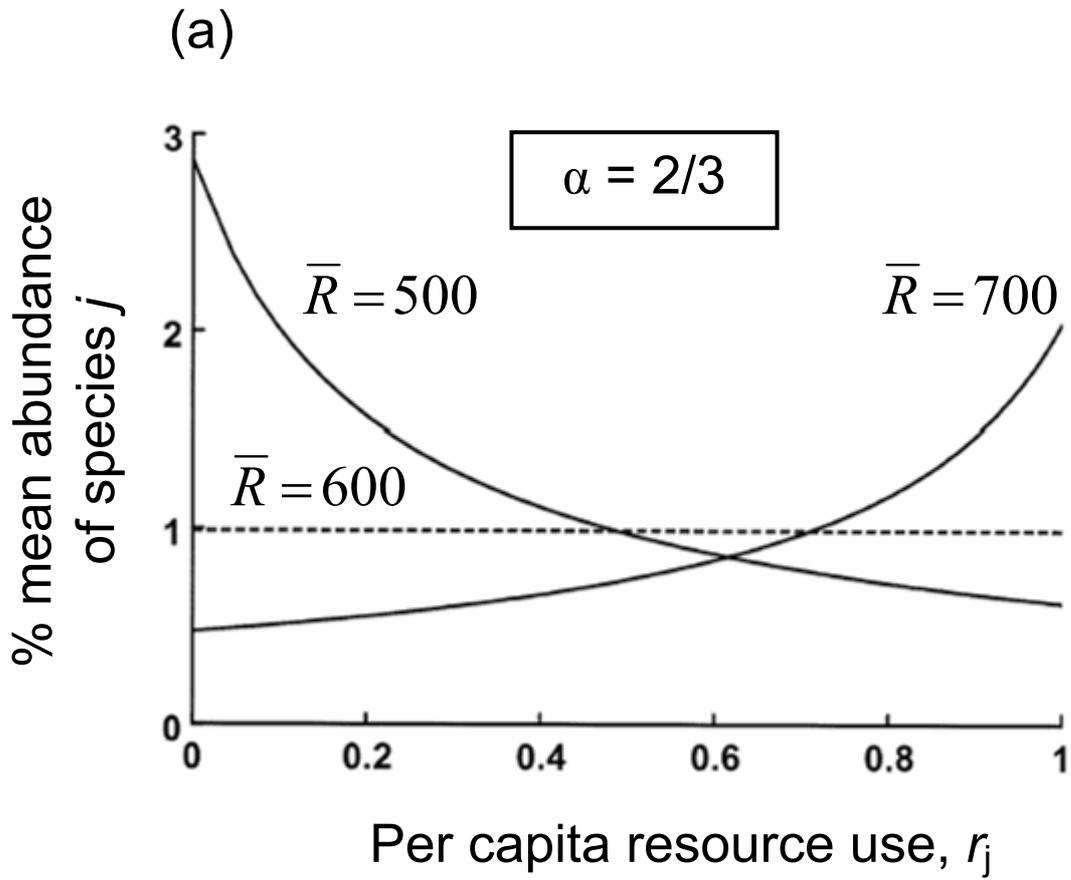

(b)

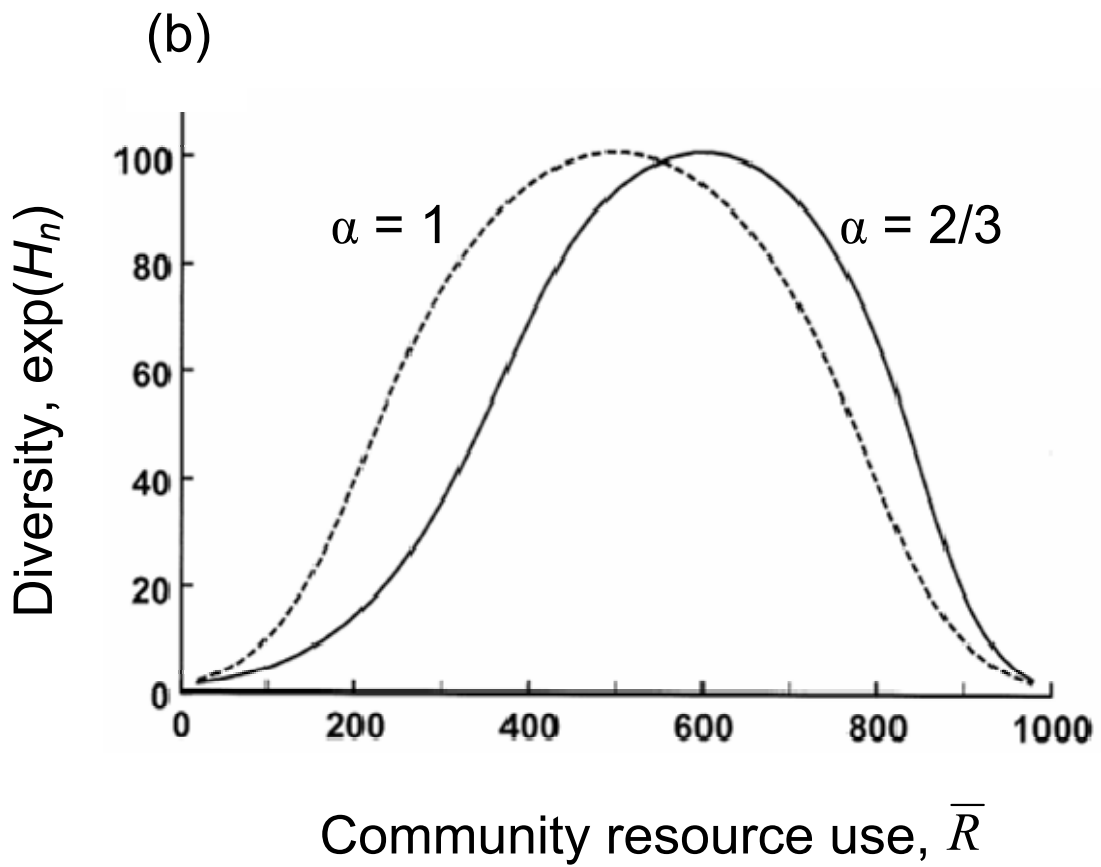

Fig. 3

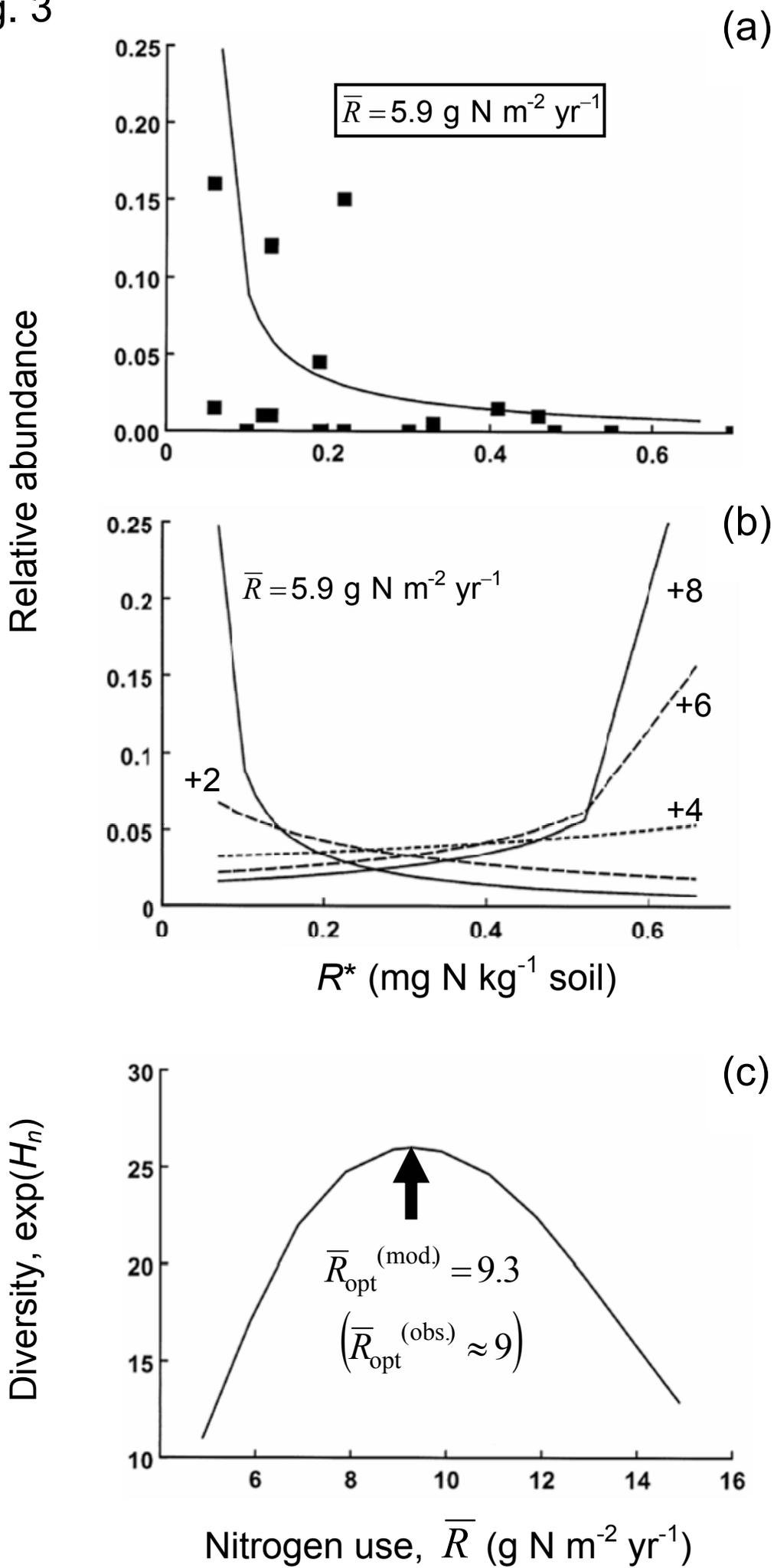

(a) $\overline{R} = 5.9$ g N m$^{-2}$ yr$^{-1}$

(b) $\overline{R} = 5.9$ g N m$^{-2}$ yr$^{-1}$

+8

+6

+4

+2

$R^*$ (mg N kg$^{-1}$ soil)

(c) $\overline{R}_{opt}^{(mod.)} = 9.3$

$\left( \overline{R}_{opt}^{(obs.)} \approx 9 \right)$

Relative abundance

Diversity, $\exp(H_n)$

Nitrogen use, $\overline{R}$ (g N m$^{-2}$ yr$^{-1}$)

Fig. 4

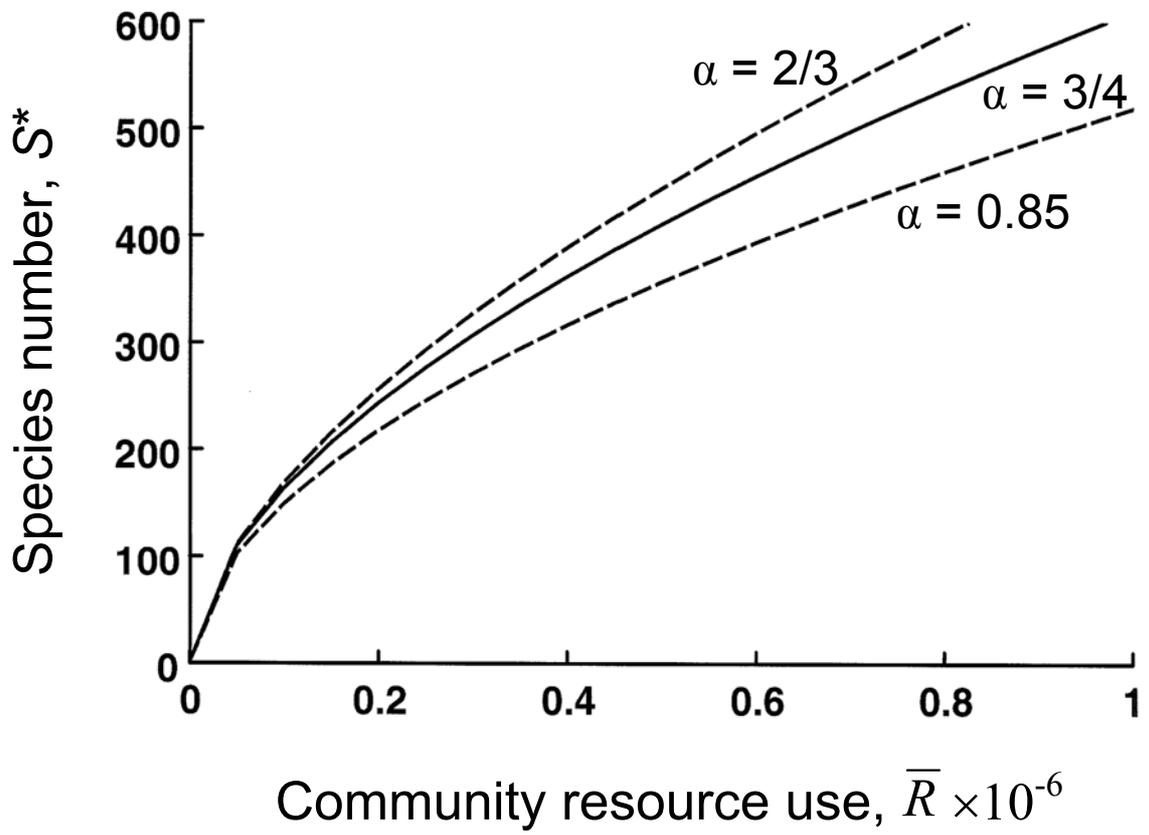